\documentstyle[12pt]{article}
\include{epsf}
\begin{document}

\begin{center}

{\bf ELECTROWEAK PHASE TRANSITION IN A STRONG MAGNETIC FIELD }\\[24pt]

{\bf \sl Vladimir Skalozub$^a$ and Michael Bordag$^b$}\\[12pt]

$^a$ Dniepropetrovsk State University, 320625 Dniepropetrovsk, Ukraine

$^b$ University of Leipzig, Augustusplatz 10,  04109 Leipzig, Germany
\end{center}                       

\noindent
ABSTRACT

Phase transitions induced by high temperatures and strong
magnetic fields are investigated in the Standard model. The consistent
effective potential including the one-loop and ring diagram
contributions is calculated and investigated for the wide range of the
fields, temperatures and Higgs boson mass $m_H$. All other particles -
fermions and bosons - are taken into account with actual values of
their masses. This effective potential is real at sufficiently high
temperatures . It is shown that symmetry restoration is a first type
phase transition for $m_H < 70 $ Gev .  For heavier Higgs particles
and the field strengths $H > 0.1 \cdot 10^{24} G $ the local
electroweak minimum could not be realized at all. Hence, the upper
limit on the Higgs mass as well as the limit on the field strengths in
the phase transition epoch follow.

    \subsection*{ 1.Introduction }

   The concept of symmetry restoration at high temperature has been
intensively used in studying the evolution of the universe at its
early stages. Nowadays it gives a possibility to investigate various
problems of cosmology and particle physics \cite{DK},\cite{ADL}. In
particular, the type of the electroweak phase transition and hence a
further evolution of the universe depends on the mass $m_{H}$ of the
Higgs boson. Most investigations of the electroweak phase transition
have included into consideration a high temperature environment as the
main ingredient \cite{ADL}, \cite{She}. But in recent a few years
cogent arguments followed from different approaches in favour of the
presence of strong magnetic fields at that stage have appeared
\cite{Vac}, \cite{Bra}( recent survey on the magnetic fields in the
universe is Ref.\cite{Enq}). So, the phase transition at high
temperature and strong fields has to be of interest . Moreover, at
present time when all masses of fundamental particles, except $m_{H}$,
are known it is possible to investigate in details the phase
transition as the function of this parameter and to determine the
properties of the vacuum, its structure and to specify a so-called
metastability bound on $m_{H}$ \cite{She},\cite{Esp}.

One of the ways to have strong magnetic fields in the electroweak
phase transition epoch was proposed by Vachaspati \cite{Vac}.  From
his analysis it follows that under very general conditions the fields
$H \sim T^2_i$ in the patches of sub-horizon scales can be generated
during a large class of grand unified transitions
\cite{Bra},\cite{Enq}, where $T_i$ is transition temperature. The
second one is formation of the Savvidy vacuum magnetic state at high
temperature ($H \sim g T^2, g $ is gauge coupling constant)
\cite{SVZ1},\cite{Enq},\cite{Sk3}. In latter case only the abelian
field configurations could arise spontaneously since they are
sourceless.  For many problems it is important to estimate the field
strengths presented, but it is difficult to realize that without
detailed investigations within specific models. Usually, only one type
of fields is considered. Therefore, results obtained in such a way
give an upper estimate of the field. This remark is relevant to our
present investigation. In current literature along with usual magnetic
field a hypercharge magnetic field presenting in the restored phase is
discussed ( see for example \cite{Shap}). The latter one is converted
into ordinary magnetic field during the phase transition.
    
In what follows, we will consider the case when the magnetic field is
presenting in both broken and restored phases. We believe that this is
a good approximation which gives possibility to investigate essential
features of the phase transition. This scenario assumes that the field
has been generated at a GUT scale via the Savvidy mechanism and
presented during the phase transition.  Such a picture is likely
since, as it follows from our consistent calculation of the effective
potential which found to be real at sufficiently high temperatures (
and, in particular, in the restored phase), constant magnetic field is
stable.  This is the most important point of the present analysis.  In
any case, it may have relevance to the description of the phase with
broken symmetry at high temperatures and strong fields.

The discussed "primordial" fields are usually considered as seed
fields responsible for generation of the observed magnetic fields in
galaxies \cite{Enq}.

Various aspects of the phase transitions in magnetic fields at high
temperature have been investigated by many authors
\cite{Cha}-\cite{SVZ2}. In Refs.  \cite{Sk1},\cite{Rez} the phase
transitions derived with the effective potential (EP) of the bosonic
part of the Salam-Weinberg model and the vacuum structures of the
phases have also been described. In Ref.\cite{SVZ2} in addition to the
temperature and magnetic fields a chemical potential was
incorporated. But the role of the fermions has not been investigated
in detail.  However, due to a rather heavy $t$-quark mass, $m_t \simeq
175$ Gev, an unbounded global minimum of the EP is produced in
addition to the electroweak local one for not very heavy Higgs scalars
\cite{She}. As will be shown, strong magnetic fields influence
essentially the phase transition dynamics realized in this
case. Another aspect of the electroweak phase transition, which also
was not investigated but plays an important role, is the influence of
so-called ring diagrams at high temperature and strong field.  At zero
field it was investigated in Refs.\cite{Tak},\cite{Car} where it has
been shown the importance of these diagrams for determining the type
of the phase transition.  In Ref.\cite{Car} the t-quark mass was
chosen of order $110 $Gev.  So, taking into account the present day
data, it should be considered as a qualitative estimate of the role of
ring diagrams even for zero-field case.

In the present paper we study the electroweak phase transition at high
temperatures and constant strong magnetic fields $H$ . We calculate
and investigate the one-loop effective potential(EP) and the
contributions of ring diagrams.  In contrast to previous
considerations we include the content of all bosons and fermions with
the corresponding masses. So, the only free parameter remains the mass
$m_{H}$. The role of ring diagrams is of great importance because they
contain the terms which cancel out the imaginary part of the one-loop
EP.  The total EP is real at sufficiently high temperatures and
suitable for investigations of symmetry behaviour.  As it will be
shown, the influence of fermions (mainly heavy quarks) is very
essential. For values of mass $m_H < 70 Gev$ and weak field strengths
the electroweak phase transition is of the first type one. It is
turned out that the magnetic field stimulates either the generation of
the electroweak minimum and tends to remove the potential barrier
separating this metastable state from the unbounded global minimum
produced due to heavy fermions. At any high temperature there exists
the corresponding field strength $H$ at which the metastable vacuum
could not be produced at all. This property can be used to find an
upper limit on $H$ at the transition epoch.  The paper is organized as
follows. In Sects.2,3 the one-loop contributions of bosons and
fermions to the EP$V^{(1)}(T,H,\phi_c)$ are calculated in the form
convenient for numerical study. In Sect.4 we compute the contributions
of ring diagrams.  Further in Sect.5 symmetry behaviour is
investigated for a number of values of $ m_{H}$ and $H$ . For
comparison, we consider separately the cases when only the one-loop EP
is taken into account and when the ring diagrams are included.
Discussion of the results obtained is given in Sect.6.

      \subsection*{2. Boson contributions to $V^{(1)}(T,H,\phi_c)$}

The Lagrangian of the boson sector of the Salam-Weinberg model is

\begin{eqnarray} \label{1}
 L = -\frac{1}{4} F^a_{\mu\nu} F^{\mu\nu}_a -\frac{1}{4} G_{\mu\nu}
G^{\mu\nu} + (D_{\mu}\Phi)^+ (D^{\mu}\Phi) \nonumber\\
+\frac{m^2}{2}(\Phi^+ \Phi) - \frac{\lambda}{4} (\Phi^+\Phi)^2,
\end{eqnarray}

where the standard notations are introduced
\begin{eqnarray}
F^a_{\mu\nu} = \partial_{\mu}A^a_{\nu}-\partial_{\nu}A^a_{\mu} +
g\varepsilon^{abc} A^b_{\mu}A^c_{\nu}, \nonumber\\ G_{\mu\nu} =
\partial_{\mu}B_{\nu}-\partial_{\nu}B_{\mu}, \nonumber\\ D_{\mu} =
\partial_{\mu} + \frac{1}{2}ig A^a_{\mu}\tau^a + \frac{1}{2}ig'
B_{\mu}.
\end{eqnarray}
\nonumber\\
The vacuum expectation value of the field $\Phi$ is
\begin{equation} \label{2} <\Phi> = \frac{1}{\sqrt{2}}\left(\begin{array}
{c}0 \\ \phi_c \end{array} \right).
\end{equation}
The fields corresponding to the $W$-,$Z$-bosons and photons, 
respectively, are
\begin{eqnarray}  W^{\pm}_{\mu} =\frac{1}{\sqrt{2}}(A^1_{\mu} \pm 
iA^2_{\mu}),
\nonumber\\
 Z_{\mu} =\frac{1}{\sqrt{g^2 + g'^2}}(gA^3_{\mu} - g'B_{\mu}),
\nonumber\\
 A_{\mu} =\frac{1}{\sqrt{g^2 + g'^2}}(g'A^3_{\mu} + gB_{\mu}).
\end{eqnarray}
\nonumber\\ The external electromagnetic field is introduced by
splitting the potential in two parts: $A_{\mu}= \bar{A_{\mu}} +
A^{R}_{\mu} $, where $A^{R}$ describes a radiation field and $\bar{A}
= (0,0,Hx^1,0)$ corresponds to the constant magnetic field directed
along the third axis. We make use of the gauge- fixing conditions
\cite{Sk2}
\begin{equation} \label{3} \partial_{\mu}W^{\pm \mu} \pm 
ie\bar{A_{\mu}}
W^{\pm \mu} \mp i\frac{g^2\phi^2}{4\xi}\phi^{\pm} = C^{\pm}(x),
\end{equation}
\begin{equation} \label{4} \partial_{\mu}Z^{\mu} - \frac{i}{\xi'}
(g^2 + g'^2)^{1/2}\phi_{z} = C_z ,
\end{equation}
where $ e = g sin \theta_w, tang \theta_w = g'/g, \phi^{\pm}, \phi_{z}$ are 
the
Goldstone fields, $\xi, \xi' $ are the
gauge fixing parameters, $C^{\pm}, C_z$ are arbitrary functions and 
$\phi_c$ is
a scalar condensate value. In what follows, all calculations will be done in
the general relativistic renormalizable gauge (\ref{3}),(\ref{4}) and after
that we  set $\xi,\xi' = 0$ choosing the unitary gauge.

To compute the EP $V^{(1)}$ in the background magnetic field let us
introduce the proper time,s, representation for the Green functions
\begin{equation} G^{ab}= - i \int\limits_{0}^{\infty} ds \exp(-is {G^{-
1}}^{ab})
\end{equation}
\nonumber\\
and use the
method of Ref.\cite{Cab}, allowing in a natural way to incorporate the 
temperature
into this formalism. A basic formula of Ref.\cite{Cab} connecting the 
Matsubara
Green functions with the Green functions at zero temperature is needed,
\begin{equation} \label{5} G^{ab}_k(x,x';T) = \sum\limits_{-
\infty}^{+\infty}
(-1)^{(n+[x])\sigma_k} G^{ab}_k(x-[x]\beta u, x'- n\beta u),
\end{equation}
where $G^{ab}_k $is the corresponding function at $T=0, \beta =1/T, u
= (0,0,0,1),$ the symbol $[x]$ means an integer part of $x_{4}/\beta,
\sigma_k = 1$ in the case of physical fermions and $\sigma_{k} =0$ for
boson and ghost fields.  The Green functions in the right-hand side of
formula (\ref{5}) are the matrix elements of the operators $G_k$
computed in the states $\mid x',a)$ at $T=0$, and in the left-hand
side the operators are averaged in the states with $T\not= 0$. The
corresponding functional spaces $U^{0}$ and $U^{T}$ are different but
in the limit of $T \ss 0$ $ U^{T}$ transforms into $U^{0}$.

The one-loop contribution into EP is given by the expression
\begin{equation} \label{6} V^{(1)} = \frac{1}{2} Tr\log G^{ab},
\end{equation}
where $G^{ab}$ stands for the propagators of all the quantum fields 
$W^{\pm},
\phi^{\pm},...$ in the background magnetic field $H$. In the s-
representation
the calculation of the trace can be done in accordance with formula 
\cite{Sch}
\begin{equation} Tr\log G^{ab} = - \int\limits_{0}^{\infty} \frac{ds}{s}
tr \exp(-is G^{-1}_{ab} )
\nonumber\\
\end{equation}
Details of calculations based on the s-representation and the formula 
(\ref{5})
can be found, for example,in Refs.\cite{Cab},\cite{Rez},\cite{Sk3}. The 
terms with $n=0$ in
Eqs.(\ref{5}),
(\ref{6}) give  zero temperature expressions for Green's functions and
effective potential $V^{(1)}$, respectively. They are the only terms 
possessing
divergences. To eliminate them and uniquely fix the potential we use the
following renormalization conditions at $H,T = 0$\cite{Rez}:
\begin{equation} \label{7} \frac{\partial^2 V(\phi,H)}{\partial 
H^2}\mid_{H=0,
\phi=\delta(0)} = \frac{1}{2},
\end{equation}
\begin{equation} \label{8} \frac{\partial V(\phi,H)}{\partial 
\phi}\mid_{H=0,
\phi=\delta(0)} = 0,
\end{equation}
\begin{equation} \label{9} \frac{\partial^2 V(\phi,H)}{\partial \phi^2}
\mid_{H=0,\phi=\delta(0)} = \mid m^2 \mid,
\end{equation}
where $V(\phi,H)=V^{(0)}+V^{(1)}+ \cdots$ is the expansion in a 
number of loops
and $\delta(0)$ is the vacuum value of a scalar field determined in  a tree
approximation.

It is convenient for what follows to introduce the dimensionless
quantities: $h=H/H_0 (H_0=M^2_w/e),\phi=\phi_c/\delta(0), K
=m_H^2/M_w^2,$ $B=\beta M_w, \tau=1/B = T/M_w,$$ {\cal V}= V/H^2_0$
and $M_w = \frac{g}{\sqrt{2}}\delta(0)$.

After reparametrization the scalar field potential is directly
expressed in terms of the ratio $K,$
\begin{equation} \label{10} {\cal V}^{(0)} = \frac{h^2}{2} + K(-
\frac{\phi^2}
{4} + \frac{\phi^4}{8} ).
\end{equation}
The renormalized one-loop EP is given by the sum of the functions
\begin{equation} \label{11}
{\cal V}_1 = {\cal V}^{(0)} + {\cal V}^{(1)}(\phi,H,K)
 + \omega^{(1)}(\phi,h,K,\tau),
\end{equation}
where ${\cal V}^{(1)}$ is the one-loop EP at $T=0$, which has been 
studied
already in Ref.\cite{Sk2}. It has the form:
\begin{equation} \label{12} {\cal V}^{(1)} = {\cal V}^{(1)}_{w,z} +
{\cal V}^{(1)}_{\phi},
\end{equation}
where
\begin{eqnarray} \label{12a} {\cal V}^{(1)}_{w,z} = 
&\frac{3\alpha}{\pi}&
[h^2 Log \Gamma_1 (\frac{1}{2} +\frac{\phi^2}{2h}) + h^2 \zeta^{'}(-1) 
+
\frac{1}{16} \phi^4 - \frac{1}{8} \phi^4 log\frac{\phi^2}{2h} + 
\frac{1}{24}
h^2 \nonumber\\&-& \frac{1}{24} h^2 log(2h)]\nonumber\\
&+& \frac{\alpha}{2\pi} [ - 2h^2 + (h^2 + h \phi^2) log(h + \phi^2) + (h^2 
-
h\phi^2 ) log \mid h - \phi^2 \mid ] \nonumber\\
&+& i \frac{1}{2} \alpha h (\phi^2 - h) \theta (h - \phi^2),
\end{eqnarray}
\begin{eqnarray} \label{12b} {\cal V}^{(1)}_{\phi} = &K& sin^2 \theta
_w ( - \phi^2 + \frac{1}{2} \phi^4 )  \nonumber\\
&+&  \frac{3 \alpha}{4\pi}( 1 + \frac{1}{2 cos^2 \theta}) (\frac{1}{2} 
\phi^4
log \phi^2 - \frac{3}{4} \phi^4 + \phi^2 )  \nonumber\\
&+& \frac{\alpha K^2}{32 \pi} [(\frac{9}{2} \phi^4 - \frac{3}{4} \phi^2 
+
\frac{1}{2} ) log \mid \frac{3 \phi^2 - 1}{2} \mid - \frac{27}{4}\phi^4 +
\frac{21}{2} \phi^2 ]
\end{eqnarray}
and $\omega^{(1)}$ is the temperature dependent contribution to the EP
determined by the terms of formulae (\ref{5}),(\ref{6}) with $n \not= 0$.

We outline the used calculation procedure considering the $W$-boson
contribution as an example \cite{Sk3},
\begin {eqnarray} \label {13}
\omega^{(1)}_{w} =  \frac{\alpha}{2\pi}
\int\limits_{0}^{\infty}\ \frac{ds}{s^2}\ e^{-is(\phi^2/h)} \Bigl[\frac{1 +
2 \cos 2s}{\sin s} \Bigr]
\sum\limits_{1}^{\infty} \exp(ihB^2 n^2/4s).
\end{eqnarray}
As Eq.(\ref{12a}), this expression contains an imaginary part for $h >
\phi^2$ appeared due to the tachyonic mode $ \varepsilon^2 = p^2_3 +
M^2_w - eH $ in the $W$-boson spectrum \cite{Sk2}. It can be
explicitly calculated by means of an analytic continuation taking into
account the shift $s \ss s -i0 $ in the $s$- plane. Fixing $\phi^2/h >
1$ one can rotate clockwise the integration contour in the $s$-plane
and direct it along the negative imaginary axis. Then, using the
identity
\begin{equation} \frac{1}{\sinh s} = 2 \sum\limits_{p=0}^{\infty} e^{-
s(2p+1)}
\end{equation}
\nonumber\\
and integrating over $s$ in accordance with the standard formula
\begin{equation} \label{14} \int\limits_{0}^{\infty} ds s^{n-1} \exp(-
\frac{b}
{s} - as) = 2 (\frac{a}{b})^{n/2} K_{n}(2\sqrt{ab}),
\end{equation}
$a,b > 0$, one can represent the expression (\ref{13}) in the form
\begin{equation} \label{14a}  Re \omega^{(1)}_w = - 4 \frac{\alpha}{\pi}
\frac{h}{B} ( 3\omega_0 + \omega_1 - \omega_2 ),
\end{equation}
where
\begin{equation} \omega_0 = \sum\limits_{p=0}^{\infty} 
\sum\limits_{n=1}^
{\infty} \frac{x_p}{n} K_1(nBx_p) ; x_p = (\phi^2 + h +2ph )^{1/2}
\end{equation}
\nonumber\\
\begin{equation} \omega_1 = \sum\limits_{n=1}^{\infty} \frac{y}{n} 
K_1(nBy),
y = (\phi^2 - h )^{1/2}
\end{equation}
\nonumber\\
and in the region of parameters $ \phi^2 < h $ after analytic continuation
\begin{equation} \omega_1 = -\frac{\pi}{2} \sum\limits_{n=1}^{\infty}
\frac{\mid y \mid}{n} Y_1(nB\mid y \mid) ,
\end{equation}
\nonumber\\
\begin{equation} \omega_2 = \sum\limits_{n=1}^{\infty} \frac{z}{n} 
K_1(nBz),
z = (\phi^2 + h )^{1/2},
\end {equation}
and $K_n(x), Y_n(x)$ are the Macdonald and Bessel functions, 
respectively. The
imaginary part of $\omega^{(1)}_w$ is given by the expression
\begin{equation} \label{15} Im \omega_2 = -
2\alpha\frac{h}{B}\sum\limits_{n=1}
^{\infty} \frac{\mid y \mid }{n} J_1(nB\mid y \mid),
\end{equation}
$J_n(x) $is Bessel function. As is well known, the imaginary part of EP is
signaling the instability of a system \cite{Sch}. In what follows we shall
consider mainly the symmetry behaviour described by the real part of the 
EP .
As  the imaginary part is concern,  it will be cancelled  in consistent 
calculation including the one-loop and ring diagram contributions to the 
EP.

Carrying out similar calculations for the $Z$- and Higgs bosons, we obtain
\cite{Sk1}:
\begin{equation} \label{16} \omega_z = - 6\frac{\alpha}{\pi} 
\sum\limits_{n=
1}^{\infty} \frac{\phi^2}{\cos^2 \theta_w n^2 B^2} 
K_2(\frac{nB\phi}{\cos\theta})
\end{equation}
\begin{equation} \label{17}
Re \omega_{\phi} = \Bigl\{
\begin{array}{c}-2 \frac{\alpha}{\pi} \sum\frac{t^2}{B^2 n^2} 
K_2(nBt)\\
\alpha \sum\limits_{n=1}^{\infty} \frac{\mid t \mid^2}{n^2 B^2} 
Y_2(nB\mid t
\mid)\end{array}\Bigr\}.
\end{equation}
where the variable $t = [K_w(\frac{3\phi^2-1)}{2})]^{1/2}$ at 
$3\phi^2>1$ and
series with the function $Y_2(x)$ has to be calculated  at $3\phi^2<1$. 
The
corresponding imaginary part is also cancelled as it will  be  shown below.

The above expressions (\ref{12}),(\ref{14a}),(\ref{16}),(\ref{17}) will be
used in numerical studying of the symmetry behaviour at different $H,T$. 
There is a cancellation of a number of terms from the zero-temperature
contributions given Eqs.(\ref{12}) and $T$-depended ones. This fact has a
general character and was used in checking of the correctness of 
calculations.

\subsection*{3. Fermion contributions to $V^{(1)}(H,T,\phi_c)$}

To find the explicit form of the fermion contribution to the EP
let us consider the standard  unrenormalized  expression  written in the $s$
-representation \cite{Elm}:
\begin{equation} \label{18} V^{(1)}_{f} = - \frac{1}{8\pi^2}\sum\limits_
{n=-\infty}^{\infty} (-1)^n \int\limits_{-\infty}^{+\infty}\frac{ds}{s^3}
e^{-(m^2s+\beta^2n^2/4s)} (eHs)coth eHs ,
\end{equation}
$m$ is a fermion mass. Here, we have incorporated the equation
(\ref{5}) to include a temperature dependence. In what follows, we
shall take into account the contributions of all fermions - leptons
and quarks - with their masses known at present time. Usually,
considering a symmetry behaviour without field one restricts himself
by a $t$-quark contribution only. But in the case of an external field
applied this is not a good idea, since the dependence of $V^{(1) }$ on
$H$ is a complicated function of the ratio $m^2_{f}/eH$. So, at some
fixed values of $H,T$ different dependencies on $H$ will contribute
for fermions with different masses. Hence, a very complicate
dependence on the field takes place in general. We include this in a
total, carrying out a numerical calculations and summing up over all
the fermions. Now, separating a zero temperature contribution by means
of the relation $\sum\limits_{-\infty}^ {+\infty} = 1 + 2
\sum\limits_{1}^{\infty}$ and introducing the parameter
$K_{f}=m^2_{f}/M^2_w = 2G^2_{Yukawa}/g^2$, we obtain for the zero
temperature fermion contribution to the dimensionless EP ,
\begin{eqnarray} \label{19}
{\cal V}_{f}(h,\phi) = &\frac{\alpha}{4\pi}& \sum\limits_{f} K^2_{f}(- 2 
\phi^2
+ \frac{3}{2} \phi^4 - \phi^4 log \phi^2 )\nonumber\\
 &-&\frac{\alpha}{\pi}\sum\limits_{f}(q^2_f\frac{h^2}{6}
\log\frac{2\mid q_f \mid h}{K_f})
\nonumber\\
&-& \frac{\alpha}{\pi} \sum\limits_{f} \Bigl.[2q^2_f h^2 \log\Gamma_1(
\frac{K_f\phi^2}{2\mid q_f\mid h}) + (2\zeta'(-1)-\frac{1}{6})q^2_f h^2
\nonumber\\
&+&\frac{1}{8} K^2_f\phi^4 + (\frac{1}{4}K^2_f\phi^4 - 
\frac{1}{2}K_f\mid q_f
\mid h\phi^2) \log\frac{2\mid q_f \mid h}{K_f \phi^2}\Bigr]
\end{eqnarray}
where $q_f$ is a fermion electric charge, the sum $\sum\limits_{f} = \sum
\limits_{f=1}^{3}(leptons) + 3 \sum\limits_{f=1}^{3}(quarks)$ counts the
contributions of leptons and quarks with their electric charges. The 
$\Gamma_1$
function is defined as follows (see, for example, survey \cite{Sk2}):
\begin{equation} \log\Gamma_1(x) = \int\limits_{0}^{x} dy 
\log\Gamma(y) +
\frac{1}{2}x(x-1) - \frac{1}{2}x\log(2\pi).
\end{equation}
\nonumber\\

The finite temperature part can be calculated in a way, described in
the previous section. In the dimensionless variables it looks as follows:
\begin{eqnarray} \label{20} \omega_{f}&=& - 4 \frac{\alpha}{\pi}\sum
\limits_{f}\Bigl\{\sum_{p=0}^{\infty}\sum_{n=1}^{\infty}(-1)^n 
\Bigl[\frac{
(2ph + K_f\phi^2)^{1/2} h}{Bn} K_1((2ph + K_f\phi^2)^{1/2}Bn)
\nonumber\\
&+& \frac{(2p+2)h + K_f\phi^2)^{1/2}}{Bn} h
K_1(((2p+2)h + K_f\phi^2)^{1/2}Bn)\Bigr]\Bigr\}
\end{eqnarray}
Again, a number of terms from Eqs.(\ref{19}) and (\ref{20}) are cancelled 
being
summed up, as in the bosonic sector.

These two expressions and the boson contributions obtained in Sect.2 will 
be
used in numerical investigations of symmetry behaviour. More precise, we
consider the difference $V'(H,T,\phi) = Re V(H,T,\phi) - Re V(H,T, 
\phi=0)$ giving
possibility to determine the symmetry restoration.  We will investigate the 
EP of two types - the one-loop contribution and the sum of that and the 
ring diagrams  which are the next to leading order corrections at high 
temperature.

\subsection*{4. Contribution of ring diagrams}

It was shown by Carrington \cite{Car} that at $T \not = 0$ a consistent 
calculation
of the EP based on generalized propagators, which include the polarization
operator insertions, requires that ring diagrams have to be added 
simultaneously with the one-loop contributions. These diagrams  essentially 
affect
the phase transition at high temperature and zero field 
\cite{Tak},\cite{Car}.
 Their importance at $ T$ and $H \not= 0$ was also pointed out in 
literature
 \cite{SVZ1},\cite{SVZ2} but, as far as we know, this part of the EP has   
not been calculated, yet. 

As is known \cite{Tak}, the sum of ring diagrams describes a dominant
contribution of great distances. It differs from zero only in the case
when massless states appear in a system. So, this type of diagrams has
to be calculated when a symmetry restoration is investigated. To find
the correction $V_{ring}(H,T)$ at high temperature and magnetic field
the polarization operators of the Higgs particle, photon and $Z$-boson
at the considered background have to be calculated. Just these
calculations have been announced in Refs.\cite{SVZ1},
\cite{SVZ2}. Then, $V_{ring}(H,T)$ is given by series depicted in
figures 1,2.
%                            Fig.1
\begin{figure}[ht]
\unitlength1cm
\begin{picture}(14,2)
\label{fd}
\put(0,-4.5){\epsfxsize=10cm\epsfysize=8cm\epsffile{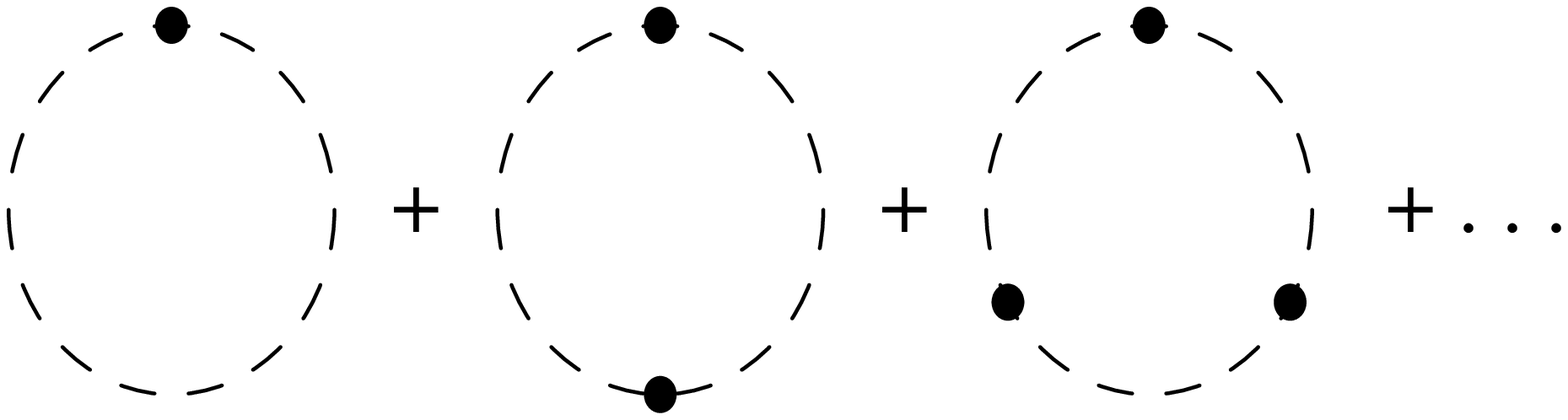}}
\end{picture}
\caption{Scalar field ring diagrams giving contribution to the effective
potential}
\end{figure}
%                            Fig.2
\begin{figure}[ht]
\unitlength1cm
\begin{picture}(14,2)
\label{fd}
\put(0,-4.5){\epsfxsize=10cm\epsfysize=8cm\epsffile{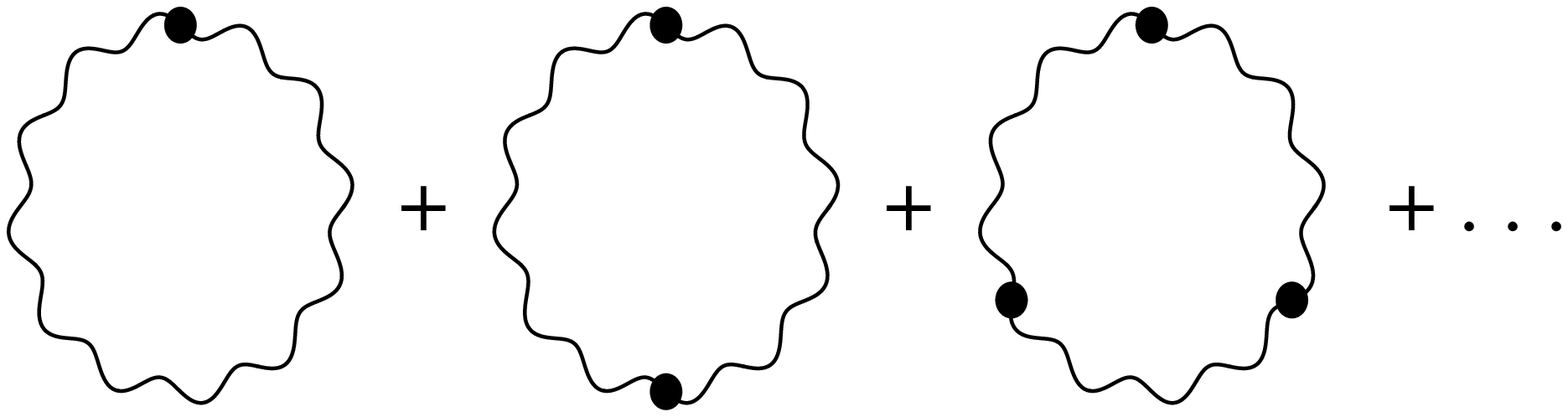}}
\end{picture}
\caption{Photon and $Z$-boson ring diagrams giving contribution to the
effective potential}
\end{figure}

Here, a dashed line describes the Higgs particles, the wavy lines
represent photons and $Z$-bosons, the blobs represent the polarization
operators in the limit of zero momenta. As also is known \cite{Car},
in order to calculate the contribution of ring diagrams not the
complete polarization operators $\Pi_{\mu\nu}(k,T,H)$ but only their
limiting expressions at zero momenta, $\Pi_{00}(k=0,T,H)$, are
sufficient. This limit, named the Debye mass, can be calculated from
the EP of the special type. This fact considerably simplifies our
task.

Now, let us turn to calculations of $V_{ring}(H,T)$. It is given by
the standard expression \cite{Tak},\cite{Car},\cite{SVZ1}:
\begin{equation} \label{21}
V_{ring} = - \frac{1}{12\pi\beta} Tr\{[M^2(\phi) +
\Pi_{00}(0)]^{3/2} - M^3(\phi)\},
\end{equation}
where trace means the summation over all the contributing states,
$M(\phi)$ is a tree mass of corresponding state and $\Pi_{00}(0) =
\Pi(k=0,T,H)$ for the Higgs particle and
$\Pi_{00}(0)=\Pi_{00}(k=0,T,H)$ are the zero-zero components of the
polarization operators in the magnetic field taken at zero momenta.
The above contribution has order $ \sim g^3 (\lambda^3) $ in coupling
constant whereas the two-loop terms are to be of order $\sim g^4 $. As
$\Pi_{00}(0)$ the high temperature limits of polarization functions
have to be substituted which have the order $\sim T^2$ for leading
terms and $\sim g\phi_c T, (gH)^{1/2}T, \phi_c, (gH)^{1/2}/T << 1$ for
subleading ones.

For the next step of calculation, we remind that the effective potential is
the generating functional of the one-particle irreducible Green functions at
zero momenta transferred. So, to have $\Pi(0)$ we can just calculate the 
second
derivative with respect to $\phi$ of the potential $V^{(1)}(H,T,\phi)$ in the
limit of high temperature ,$T >>\phi,(eH)^{1/2}$ and then set $\phi = 0$. 
This limit can be calculated by means of the Mellin
transformation technique (see for example \cite{Sk3}) and the result looks 
as
follows:
\begin{eqnarray} \label{22} V^{(1)}(H,\phi,T \ss \infty) &=& \left.[\Bigl(
\frac{C_f}{12}\phi^2_c + \frac{\alpha\pi}{2 cos^2\theta_w}\phi^2_c + 
\frac{g^2}
{16}\phi^2_c \Bigr) T^2 \right.]
\nonumber\\
&+& \left.[ \frac{\alpha \pi}{6} (3\lambda\phi^2_c - \delta^2(0))T^2 - 
\frac{
\alpha}{cos^3\theta} \phi^3 T \right.\nonumber\\
&-& \frac{\alpha}{3} (\frac{3\lambda\phi^2_c -
\delta^2(0)}{2})^{3/2} T ]
\nonumber\\
&-& \frac{1}{2\pi} (\frac{1}{4}\phi^2_c + gH)^{3/2} T + \frac{1}{4\pi} 
eH T
(\frac{1}{4}\phi^2_c + eH )^{1/2} \nonumber\\
&+& \frac{1}{2\pi} eH T (\frac{1}{4} 
\phi^2_c -
eH )^{1/2}.
\end{eqnarray}
The parameter $C_f = \sum\limits_{i=1}^{3} G^2_{il} +
3\sum\limits_{i=1}^{3} G^2_{iq}$ determines the fermion contribution
of the order $\sim T^2$ having relevance to our problem. We also
omitted $\sim T^4$ contributions to the EP.  The terms of the type
$\sim log [T/f(\phi,H)]$ cancel the logarithmic terms in the
temperature independent contributions (\ref{11}),(\ref{18}).
Considering the high temperature limit we restrict ourselves by linear
and quadratic in $T$ terms, only.

Now, differentiating this expression twice with respect to $\phi$ and
 setting $\phi=0$, we obtain
\begin{eqnarray} \label{23} \Pi_{\phi}(0) &=& \frac{\partial^2 V^{(1)}
(\phi,H,T)}{\partial \phi^2} \mid_{\phi=0}
\nonumber\\
&=& \frac{1}{24\beta^2}\Bigl( 6\lambda + \frac{6 e^2}{\sin^2 
2\theta_w}
+ \frac{3 e^2}{\sin^2 \theta_w} \Bigr) + \sum\limits_{f} 2 G^2_f/\beta^2
\nonumber\\
&+& \frac{(eH)^{1/2}}{8\pi \sin^2\theta_{w}\beta} e^2 (3\sqrt{2} \zeta(-
\frac{1}
{2},\frac{1}{2}) - 1 ).
\end{eqnarray}
The terms $\sim T^2$ in Eq.(\ref{23}) give standard contributions to
temperature mass squared coming from the boson and fermion sectors.
The $H$-dependent term is negative since the difference in the brackets is
$3\sqrt{2}\zeta (-\frac{1}{2},\frac{1}{2}) - 1 \simeq - 0,39$. Formally, 
this
may result in the negativeness of the $\Pi(0)_{\phi}$ for very strong fields
$(eH)^{1/2} > T $. But this happens in the range of parameters where 
asymptotic
expansion is not valid. Substituting expression (\ref{23}) into Eq.(\ref{21}) 
we  obtain (in the dimensionless variables),
\begin{equation} \label{24} {\cal V}^{\phi}_{ring} = - \frac{1}{12B}
\Bigl\{(\frac{3\phi^2 - 1}{2} K  + \Pi_{\phi}(0) \Bigr\}^{3/2} + 
\frac{\alpha}
{3B} K (\frac{3 \phi^2 - 1}{2})^{3/2}.
\end{equation}
As is seen, the last term of this expression cancels  the  fourth term in 
Eq.(\ref
{22}), which becomes imaginary at $3\phi^2 < 1$. This is the  important
 cancellation preventing  the infrared instability  at high temperature.

Before to proceed, let us note that Eq.(\ref{22}) contains other term
(the last one) which becomes imaginary for strong magnetic fields or
small $\phi^2$.  It reflects the known instability in the $W$-boson
spectrum which is discussed for many years in literature (see papers
\cite{Rez},\cite{SVZ1},\cite{SVZ2}, \cite{Sk3} and references
therein).  But it also will be cancelled out when the contribution of
ring diagrams with the unstable mode is added.
  
To find the $H$-dependent Debye masses of photons and $Z$-bosons the
following procedure will be used. We calculate the one-loop EP of the
$W$-bosons and fermions in a magnetic field and some "chemical
potential",$\mu$, which plays the role of an auxiliary parameter.
Then, by differentiating them twice with respect to $\mu$ and setting
$\mu = 0$ the mass squared $m^2_D$ will be obtained. Let us
demonstrate that in more detail for the case of fermion contributions
where the result is known.

The temperature dependent part of the one-loop EP of constant magnetic
field and a non-zero chemical potential induced by an
electron-positron vacuum polarization is \cite{Elm}:
\begin{equation} \label{25} V^{(1)}_{ferm.} =
 - \frac{1}{4\pi^2} \sum\limits_{l-1}^{\infty}
(-1)^{l+1}\int\limits_{0}^{\infty} \frac{ds}{s^3} exp(\frac{-\beta^2 l^2}
{4s} - m^2s ) eHs coth(eHs) cosh(\beta l\mu),
\end{equation}
where $m$ is the electron mass, $e = g sin \theta_w$ is electric
charge and a proper-time representation is used. Its second derivative
with respect to $\mu$ taken at $\mu = 0$ can be written in the form,
\begin{equation} \label{26}  \frac{\partial^2 V^{(1)}_{ferm.}}{\partial 
\mu^2}=
\frac{eH}{\pi^2}\beta^2 \frac{\partial}{\beta^2}\sum\limits_{l=1}^{\infty}
(-1)^{l+1}\int\limits_{0}^{\infty} \frac{ds}{s} exp(-m^2s -\beta^2 l^2/4s)
coth(eHs).
\end{equation}
Expanding $coth (eHs) $ in series and integrating over $s$ in accordance 
with
formula (\ref{14}) we obtain in the limit of $ T >> m, (eH)^{1/2}$:
\begin{equation} \label{27} \sum\limits_{l=1}^{\infty} (-1)^{l+1} 
[\frac{8m}
{\beta l} K_1(m\beta l) + \frac{2}{3}\frac{(eH)^2 l\beta}{m} K_1(m\beta 
l)+
\cdots ]
\end{equation}
Series in $l$ can easily be calculated by means of the Mellin transformation
(see Refs.\cite{Sk3},\cite{SVZ2}). To have the Debye mass squared it is 
necessary
to differentiate Eq.(\ref{26}) with respect to $\beta^2$ and to take into
account the relation of  the parameter $\mu$ with the zero component of 
the
electromagnetic potential : $\mu \ss ieA_0$ \cite{SVZ1}. In this way we 
obtain
finally,
\begin{equation} \label{28} m^2_{D} = g^2 sin^2\theta_w ( 
\frac{T^2}{3} -
\frac{1}{2\pi^2} m^2 + O((m\beta)^2, (eH\beta^2) ) ).
\end{equation}
This is the well known result calculated from the photon polarization
operator \cite{VZM}. As one can see, the dependence on $ H$ appears in
the order $\sim T^{-2}$. To find the total fermion contribution to
$m^2_D$ one should sum up expression (\ref{28}) over all fermions and
substitute their electric charges.

To calculate $m^2_D$ for $Z$-bosons it is sufficient to take into
account the fermion couplings with $Z$-field. It can be done by
substituting $ \mu \ss i(g/2 cos \theta_w + g sin^2 \theta_w ) $ and
the result differs from Eq.  (\ref{28}) by the coefficient at the
bracket in the right-hand side which have to be replaced, $g^2 sin^2
\theta \ss g^2 (\frac{1}{4 cos^2 \theta_w} + tang^2 \theta_w) $. One
also should add the terms coming due to the neutral currents and the
part of fermion-Z-boson interaction which is not reproduced by the
above substitution:
\begin{equation} m^{2'}_{D} = \frac{g^2}{8 cos^2 \theta_w} (1 + 4 sin^4
\theta_w) T^2 .
\end{equation}
\nonumber\\

Now, let us apply this procedure for the case of the $W$-boson 
contribution.
The corresponding EP (temperature dependent part) calculated at non-zero 
$T,\mu
$ in the unitary gauge looks as follows,
\begin{equation} \label{29} V^{(1)}_w = - 
\frac{eH}{8\pi^2}\sum\limits_{l=1}^{
\infty}\int\limits_{0}^{\infty} \frac{ds}{s^2}exp(-m^2 s -
l^2\beta^2/4s)[\frac{
3}{sinh(eHs)}+ 4 sinh(eHs)] cosh(\beta l\mu).
\end{equation}
All the notations are obvious. The first term in the squared brackets
gives the triple contribution of the charged scalar field and the
second one is due to the interaction with a $W$-boson magnetic
moment. Again, after differentiation twice with respect to $\mu$ and
setting $\mu = 0$ it can be written as
\begin{equation} \label{30} \frac{\partial^2 V^{(1)}_w}{\partial \mu^2} =
\frac{eH}{2\pi^2}\beta^2\frac{\partial}{\partial\beta^2} 
\sum\limits_{l=1}^
{\infty}\int\limits_{0}^{\infty}\frac{ds}{
s}exp(- \frac{m^2s}{eH} - \frac{l^2\beta^2eH}{4s})[\frac{3}{sinh(eHs)} 
+ 4
sinh(eHs)].
\end{equation}
Expanding $sinh^{-1}s$ in series over Bernoulli's polynomials,
\begin{equation} \frac{1}{sinh s} = \frac{e^{-s}}{s} 
\sum\limits_{k=0}^{\infty}
\frac{B_k}{k!}(-2s)^k,
\end{equation}
\nonumber\\
and carrying out all the calculations described above, we obtain for the 
$W$-
boson contribution to $m^2_D$ of the electromagnetic field,
\begin{eqnarray} \label{31} m^2_D &=& 3 g^2 sin^2 \theta_w [ \frac{1}{3} 
T^2 -
\frac{1}{2\pi} T(m^2 + g sin\theta_w H)^{1/2} - \frac{1}{8\pi^2}(g 
sin\theta_w H)\nonumber\\
&&+ O(m^2/T^2, (g sin\theta_w H /T^2)^2)].
\end{eqnarray}
Hence it follows that  spin does not contribute to the Debye mass in the 
leading order. Other interesting point is that the next to leading terms are 
negative. 

The contribution of the $W$-boson sector to the $Z$-boson mass 
$m^2_D$ is given
by expression (\ref{31}) with the replacement $g^2 sin^2 \theta_w \ss g^2 
cos^2
\theta_w$.

Summing up the expressions (\ref{28}) and (\ref{31}) and substituting
them in Eq.(\ref{21}), we obtain the photonic part $V^{\gamma}_{ring}$
where it is necessary to express masses in terms of the vacuum value
of the scalar condensate $\phi_{c}$. In the same way the contribution
of $Z$-bosons $V^{z}_{ ring}$ can be calculated. The only difference
is an additional mass term of $Z$ -field and the additional term in
the Debye mass due to the neutral current $\sim
\bar{\nu}\gamma_{\mu}\nu Z_{mu}$ . These three fields - $\phi, \gamma,
Z$ ,- which becomes massless in the restored phase, contribute into
$V_{ring}(H,T)$ in the presence of the magnetic field. At zero field
there is also a term due to the $W$-boson loops in Figs.1,2 . But when
$H \not = 0$ the charged particles acquire $\sim eH$ masses. The
corresponding fields remain short-range ones in the restored phase of
the vacuum and therefore do not contribute.

A separate consideration should be spared to the tachyonic (unstable)
mode in the $W$-boson spectrum: $p^2_0 = p^2_3 + M^2_w - eH$.  First
of all we note that this mode is produced due to a spin interaction
and it does not influence the $G_{00}(k)$ component of the $W$-boson
propagator.  Secondly, in the fields $eH \sim M^2_w$ the mode becomes
a long range state. Therefore it should be included in $V_{ring}(H,T)$
side by side with other considered neutral fields.  But in this case
it is impossible to take advantage of formula (\ref{21}) and one has
to return to the initial EP with generalized propagators .

For our purpose  it will be convenient to use the expression for the  
generalized EP written as a sum over the modes in external magnetic field 
\cite{SVZ1},\cite{SVZ2}:
\begin{equation} \label{TDVZ}  V^{(1)}_{gen} =  \frac{eH}{2\pi \beta} 
\sum \limits_{l= -\infty}^{+ \infty} \int\limits_{- \infty}^{+ \infty}
\frac{dp_3}{2\pi} \sum\limits_{n = 0, \sigma = 0,\pm 1}^{\infty} log
[\beta^2(\omega^2_l + \epsilon ^2_{n,\sigma,p_3} + \Pi(T,H) )] ,
\end{equation}
where $\omega_l = \frac{2\pi l}{\beta}$ ,  $\epsilon^2_n = p_3^2 + 
M^2_w + (2n + 1 - 2\sigma) eH $ and $\Pi(H,T)$ is the Debye mass of  
$W$-bosons in a magnetic field. 
Denoting   as $D^{- 1}_0(p_3,H.T)$ the sum $ \omega^2_l + \epsilon^2$, 
one can rewrite eq. (\ref{TDVZ}) as follows:
\begin{eqnarray} \label{TDVZ1} V^{(1)}_{gen} &= &
\frac{eH}{2\pi\beta}
\sum\limits_{-\infty}^{+\infty}\int\limits_{-\infty}^{+\infty} \frac{d
p_3}{2\pi} \sum\limits_{n,\sigma} log[\beta^2 D^{- 1}_0(p_3,H,T)]
\nonumber\\
&+&  \frac{eH}{2\pi\beta} \sum\limits_{- \infty}^{+ \infty} \int\limits_{-
infty}^{+\infty} \frac{d p_3}{2\pi} \{ log[ 1 + ( \omega^2_l + p^2_3 +
M^2_w - eH)^{-1} \Pi(H,T) ] \nonumber\\
&+&  \sum\limits_{n \not = 0, \sigma \not = +1 }
log[ 1 +  D_0 ( \epsilon^2_n , H,T) \Pi(H,T) ] \}.
\end{eqnarray}
Here, the first term is just the one-loop contribution of $W$-bosons,
the second one gives the sum of ring diagrams of the unstable mode (
as it can easily be verified by expanding the logarithm into a series
).The last term describes the sum of the short range modes and should
be omitted.

Thus, to find $V^{unstable}_{ring}$ one has to calculate the second
term in Eq. (\ref{TDVZ1}). In the high temperature limit we obtain:
\begin{equation} \label {Vunst} V^{unstable}_{ring} = 
\frac{eH}{2\pi\beta} \{ ( M^2_w - eH + \Pi (H, T) )^{1/2} - ( M^2_w - eH
)^{1/2} \}.
\end{equation}
By summing up  the one-loop EP and  all the  terms $ V_{ring}$ , we 
arrive at the total consistent in leading order effective potential. 

Let us note the most important features of the above expression. It is
seen that the last term in Eq.(\ref{Vunst}) exactly cancels the
«dangerous» term in Eq.(\ref{22}). So, no instabilities appear
at sufficiently high temperatures when $\Pi ( H,T) > M^2_w - eH $ and
the EP is real.To make a quantitative estimate of the range of
validity of the total EP it is necessary to calculate the mass
operator of $W$-boson in a magnetic field at finite temperature and
hence to find $ \Pi(H,T)$ . This is a separate and enough complicated
problem which will not be solved here completely.  Instead that below
we restrict ourselves by the contribution to $\Pi (T)$ of the neutral
Higgs particles only which can easily be calculated to give
$\Pi(T)_{Higgs} = \frac{1}{12} g^2 T^2 $.  Since other particles have
also to contribute into the temperature mass in leading $T^2$ order,
the obtained mass to be lower value which can be substituted into
$V^{unstable}_{ring}$.  Just this value will be used in the following
estimations.

\subsection*{5. Symmetry behaviour in a magnetic field at high 
temperature}

Having obtained the one-loop EP described by formulae
(\ref{12}),(\ref{14a}), (\ref{16})-(\ref{20}) and the ring diagram
contributions $V_{ring}$ we are going to investigate symmetry
behaviour at high temperature and strong magnetic fields. We shall
present the results in two steps. First, we consider the sum of the
tree and one-loop effective potentials as the function of $\phi^2$ at
various fixed $H$, $T$ and $K$ . Then, we shall add the term
$V_{ring}$ and calculate symmetry behaviour for the total EP at the
same fixed $H,T,K$. This will help to clarify the role of the plasmon
contributions. Since constructed EP includes as an input all
fundamental particles, we shall obtain new information about the
electroweak phase transition. The one-loop EP contains the imaginary
parts in some domains of $\phi^2$. But these terms occur to be
cancelled by the corresponding ones from $V_{ring}$. Thus, only the
real part,$ Re V^{(1)}$, is of interest.

As usually \cite{Sk2}, to investigate symmetry behaviour let us
consider the difference ${\cal V^{'}} = Re [{\cal V}(h,\phi, K, B) -
{\cal V}(h, \phi= 0, B )]$ which gives information about the symmetry
restoration.  Below, we consider the case when the mass $m_{H}$ is
equal to $M_w$. Typical curves for small fields $h$ and different
values of $B $ are plotted in Fig.3.

It is seen that the well known symmetry restoration (for the heavy fermion 
case) takes place. There are two minima - local, produces due to the Higgs 
mechanism, and and global one, generated by  heavy fermion contributions. 
At low temperatures (big $ B$), the local metastable minimum is disposed 
near the value $\phi^2 = 1$ that corresponds to the spontaneously broken 
symmetry. With a temperature increasing the local minimum becomes 
shallower and at $ B  \sim 0.1 $ removes to the value $\phi^2 = 0 $ that 
signals the symmetry restoration.
We see that typical second type phase transition takes place for $K = 1$.
At the same time,  the barrier separating two minima is increasing in height 
and width and  so  a  tunneling to the global unbounded from below 
minimum is suppressed.

In Fig.4 we present the influence of the field on the symmetry
behaviour at low temperature. As is seen, an increase in $h$ leads to
the getting deeper of the local minimum and to the growing up the
barrier which separates the minima. In this way the magnetic field
prevents tunneling to the global unbounded minimum.

Now, let us investigate symmetry behaviour at high temperature and
strong magnetic fields.  The result of calculations is shown in Fig.5.
From the plot it follows that the field tends to decrease the
temperature and stimulates the symmetry breaking. First the metastable
vacuum is generated with the field increasing. As is seen, this is a
homogeneous transition. When the strength $h$ is growing further the
potential barrier separating the local and global vacua is diminishing
and the depth of the former one is getting shallower. At $h \geq 2 $
the electroweak minimum disappears at all. This picture is typical and
realized at any high temperature. Therefore, it is possible to obtain
an upper limit on the magnetic field strength requiring that the
electroweak vacuum must be a long living state. From the above
analysis one could conclude that the fields $H \geq 2 H_0 = M^2/e \sim
2 \cdot 10^{24} G$ had not been generated in the early universe. In
the opposite case our world would never been realized and the system
from the very beginning suppose to be in the global unbounded
minimum. Similar symmetry behaviour (with slightly different values of
$h, B$) has also been determined for $K = 2$.

Fore completeness, let us describe the symmetry behaviour for the EP
without the fermion contributions. In this case the symmetry
restoration ( for $ K = 1,2$) is also realized by the second type
phase transition but typical temperatures to be of order $B \sim 0.5 -
0.6$.  Naturally, no global minimum exists, so no limits on the
magnetic field strength can be derived.

 To summarize the above results we stress that fermions affect in a
very essential way the symmetry behaviour in the field $H$. We also
recall that our consideration was based on the one-loop EP only.

Now, let us include in our consideration the contribution of $
V_{ring}$ .  In Fig. 6 we show the plot of $V_{ring}$ .  It represents
a complicated dependence on $h $ and this contribution for strong
fields acts to remove the separation barrier and stimulate the
transition to the global vacuum state.

In Figs.7 ,8 and 9, 10 the influence of the ring diagrams is
represented for small $ B $ , weak fields $h \sim 0.01 - 0.1$ and $K =
0.5 , 0.75$ ,respectively. To better clarify their role we show the
plots in parallel for chosen $K$.  As is seen, with $V_{ring}$
included symmetry behaviour is considerably changed as compare with
results presented in Fig. 5. Most important fact is that For $K = 1 $
the local minimum is not realized at all even for weak field
strengths. Actually, the value of $K =0.75$ corresponding $m_H \sim 70
$Gev is a bound value for the mass $m_H$ and the field strengths $H
\sim 0.01 - 0.1 M^2_w/e = 0.01 - 0.1 \cdot 10^{24}$ G give upper
limits on the magnetic fields in the phase transition epoch. Stronger
fields stimulates straight transition to the global minimum.Thus we
see that due to $V_{ring}$ term the upper limit on the magnetic field
is decreased from $2 \cdot 10^{24} G$ derived with the EP $V^{(1)}$ to
$0.1 \cdot 10^{24} G$. Moreover, they restrict the Higgs boson mass:
$m_H < 70 Gev$, otherwise the local electroweak minimum is not
produced. We also observe that the phase transition is to be of the
first type one.

It is very important that the described minima of the EP are stable at
high temperatures even for strong fields $h$. Really, typical value of
$B$ when the symmetry restoration happens for $K \sim 1$ is $ B \sim
0.1$ .  For the calculated lower temperature mass we have $
\frac{1}{12} g^2/B^2 \sim 3 $. Hence we find that the effective mass
of the unstable mode $ \phi^2 - h + \frac{1}{12} g^2/B^2 $ is positive
for all values of $h$ considered. Thus, one has to conclude that
classical constant magnetic field must be stable in the electroweak
phase transition epoch. No $W$- and $Z$-boson condensates can be
produced at high temperatures. These condensates would be realized at
lower temperatures as the intermediate states of the vacuum.

\subsection*{6. Discussion}

Let us discuss the vacuum stability in a magnetic field. This problem
is of interest because of the presence in the $W$-boson spectrum the
mode $p^2_0 = M^2_w - eH$ which becomes unstable for $H >
M^2_w/e$. The evolution of this state and its consequences have been
investigated in various aspects by many authors
\cite{AO2},\cite{Sk2},\cite{AO}.  At high temperatures, it was studied
for the case when only the bosonic part of the electroweak theory has
been included \cite{Sk1},\cite{Rez},\cite{AO}. In Refs. \cite{AO} the
classical equations of $W,Z$ and $\phi$ fields were solved and the
results that at high temperature the symmetry is restored and the
magnetic field is screened by the $W$- boson and $Z$-boson condensates
have been elaborated.  The key point in this analysis is the
assumption that symmetry restoration is the second type phase
transition which can be taken into account by adding the term $\sim
\phi^2T^2 $ in the field equations. For these results it was also
important that the Higgs boson mass equals to $m_Z$.  From the point
of view of the present investigation the described approach is not
trusty because it does not take into account the vacuum polarization
which is very important at high temperature and strong fields.  In
fact, this is the only one term produced by the vacuum polarization
and other relevant contributions must be included.  Exactly the
polarization effects produce the temperature masses which stabilize
the vacuum. So, no conditions for generation of the $W$- and $Z$-boson
condensates exist. These condensates could be realized in strong
magnetic fields at intermediate temperatures when imaginary part of
the EP is nonzero. But in any case, to have a consistent picture the
vacuum polarization should be taken into account because of a
complicated behaviour of the EP in an intermediate range of $T,H$
\cite{Sk1},\cite{Rez}.

The present investigation carried out on the base of the complete EP
with fermions included shown that at «weak» magnetic fields the
symmetry restoration is to be the first type phase transition ( for
the values of $K$ considered ,$ m_H < M_w$ ).  For strong fields our
metastable vacuum could never be realized.  But at any fixed
temperature $T$ there exists a corresponding magnetic field strength
at which a metastable vacuum with positive energy and non-zero scalar
field is produced . This picture may be of interest for cosmology.

As follows from the results of Sect.5 , the role of fermions is very
essential in the symmetry dynamics. Actually, their contribution
determines the properties of the phase transition due to magnetic
field. We have seen that at low temperature the field acts to prevent
the phase transition from the metastable to stable minimum of the
EP. Since the charged fermions and gauge bosons oppositely influence
the symmetry behaviour in a magnetic field, for the actual values of
particle masses these two contributions compensate each other and the
metastable minimum position remains near the initial point $\phi^2 = 1
$ for any values of $h$. The influence of the field is expressed in
the change of the potential barrier separating two minima. As is also
occurred, at high temperatures the role of the ring diagrams is
important (as also takes place at $h = 0$ \cite{Tak},\cite{Car}).
Thus, we conclude that the EW phase transition in a magnetic field
acquires substantial changes. For its detailed investigation it is
necessary to calculate the bubble nucleation parameters, the
metastability bound on $m_H$, etc.

We would like to complete our discussion with a few remarks concerning 
the magnetic field in the restored phase. As was mentioned in the 
Introduction, in current literature a hypercharge magnetic field is discussed 
as a relevant one there \cite{Shap}. There is an important difference 
between this field and ordinary magnetic field investigated in the present 
paper. The former field is an abelian one and requires an external source to 
maintain it in space. On the contrary, ordinary magnetic field may be 
considered as the projection of a nonabelian field produced spontaneously 
at high temperatures via Savvidy's mechanism.  Investigation of the state  
was given in Refs.\cite{Enq},\cite{SVZ1},\cite{Sk3}.  A final conclusion 
about this phenomenon can be done when the complete temperature mass 
of the unstable mode $\Pi(H,T)$ will be calculated. This work is in
progress now.

One of the authors (V.S.) expresses kind gratitude to colleagues from
Institute of Theoretical physics University of Leipzig for hospitality
and the DAAD programme for financial support of his stay at Leipzig in
period when this work had been done. The authors thank Vadim Demchik
for halping in numerical calculations and preparation of figures.

%                            Fig.3
\begin{figure}[ht]
\unitlength1cm
\begin{picture}(14,7)
\label{fd}
\put(0,-0){\epsfxsize=10cm\epsfysize=8cm\epsffile{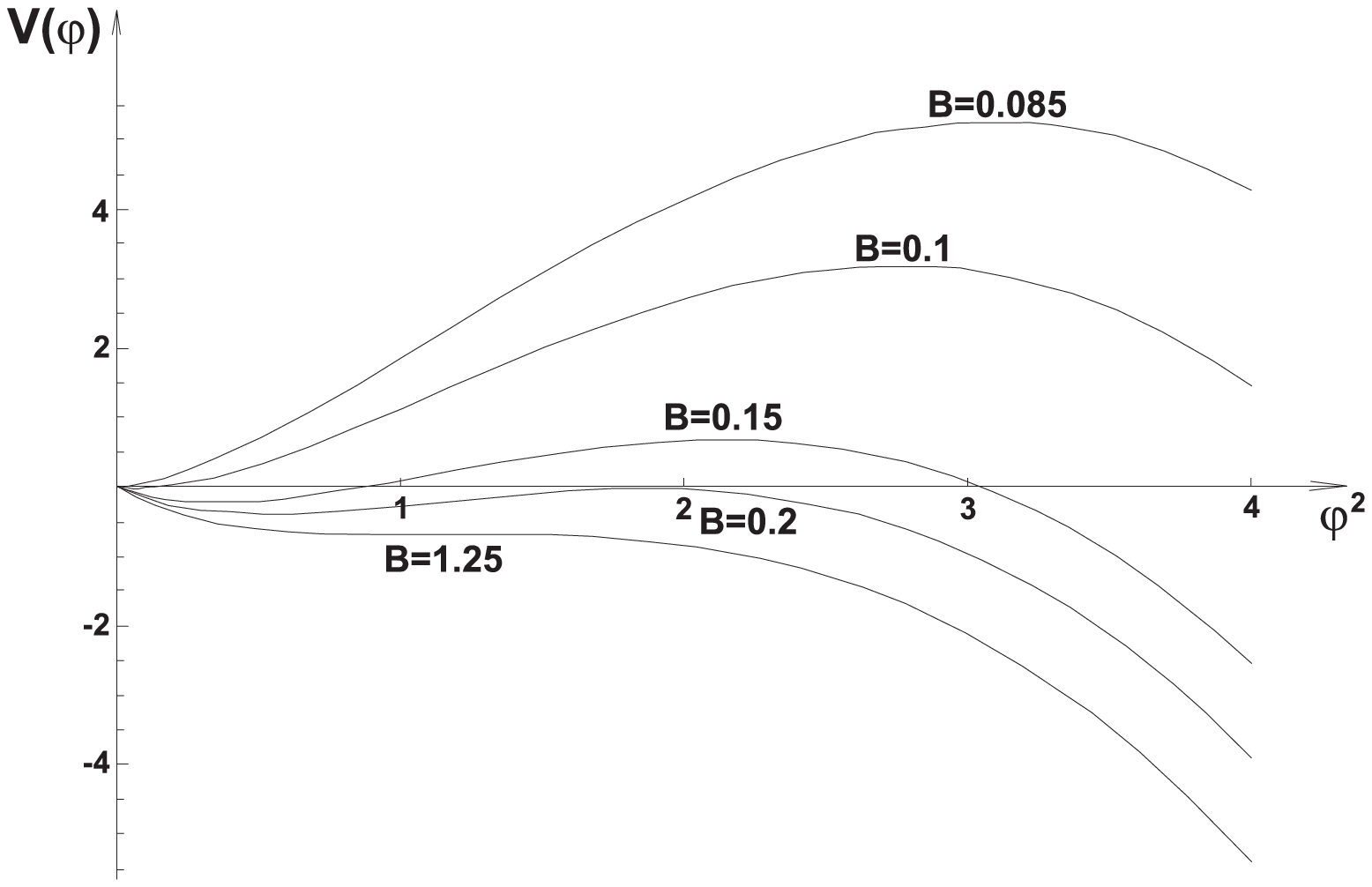}}
\end{picture}
\caption{Symmetry restoration at high temperatures and small magnetic
fields determined by  the one-loop effective potential ${\cal
V^{'}}(\phi^2, h,B,K)$.}
\end{figure}
%                            Fig.4
\begin{figure}[ht]
\unitlength1cm
\begin{picture}(14,7)
\label{fd}
\put(0,-0){\epsfxsize=10cm\epsfysize=8cm\epsffile{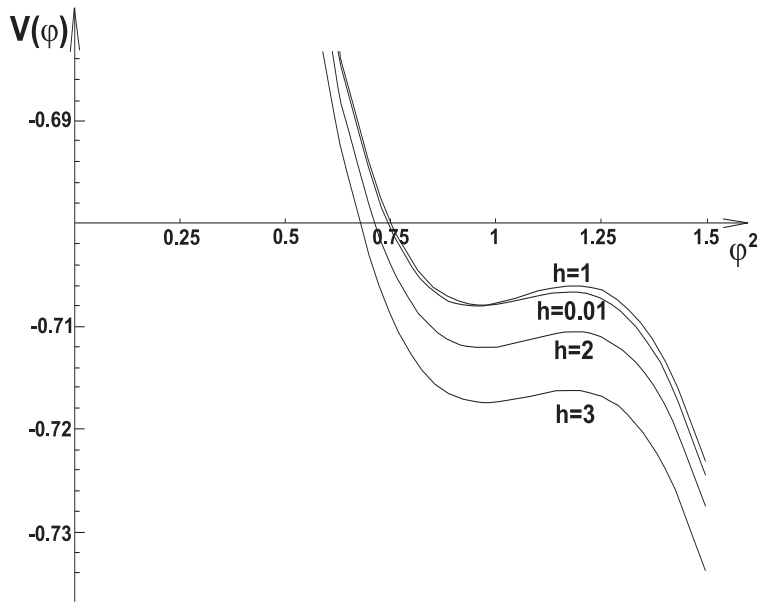}}
\end{picture}
\caption{Symmetry behaviour at zero temperature and strong magnetic
fields determined by  the one-loop effective potential ${\cal
V^{'}}(\phi^2, h,K)$.}
\end{figure}
%                            Fig.5
\begin{figure}[hb]
\unitlength1cm
\begin{picture}(14,7)
\label{fd}
\put(0,-0){\epsfxsize=10cm\epsfysize=8cm\epsffile{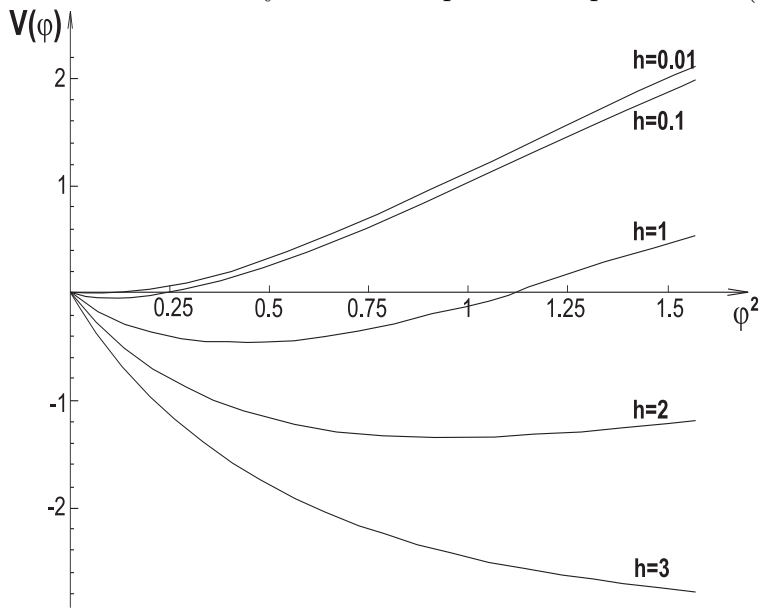}}
\end{picture}
\caption{Symmetry behaviour at fixed high temperature and a number  of
values  $h$ described by the one-loop efective potential ${\cal
V^{'}}(\phi^2,h,B,K)$.}
\end{figure}
%                            Fig.6
\begin{figure}[hb]
\unitlength1cm
\begin{picture}(14,7)
\label{fd}
\put(0,-0){\epsfxsize=10cm\epsfysize=8cm\epsffile{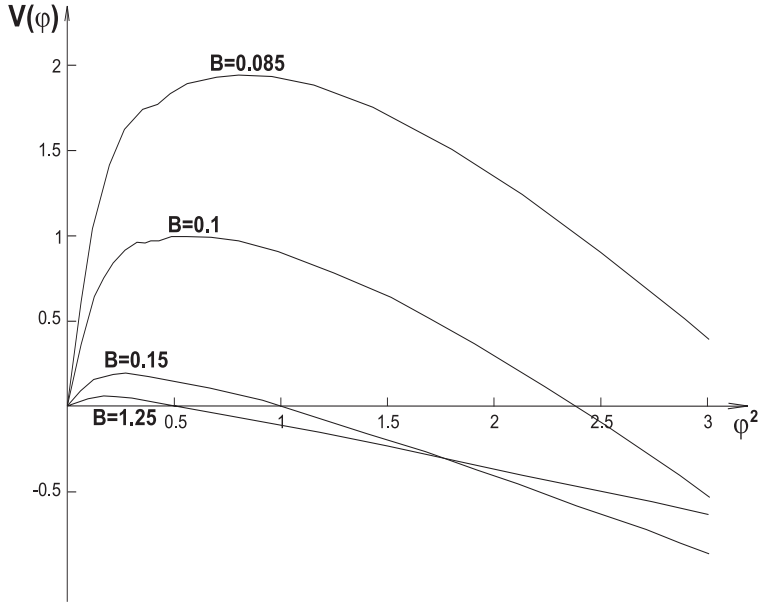}}
\end{picture}
\caption{The $V_{ring} $ curves as the functions of $\phi^2$ for  fixed
temperature and a number of $h$.}
\end{figure}
%                            Fig.7
\begin{figure}[hb]
\unitlength1cm
\begin{picture}(14,7)
\label{fd}
\put(0,-0){\epsfxsize=10cm\epsfysize=8cm\epsffile{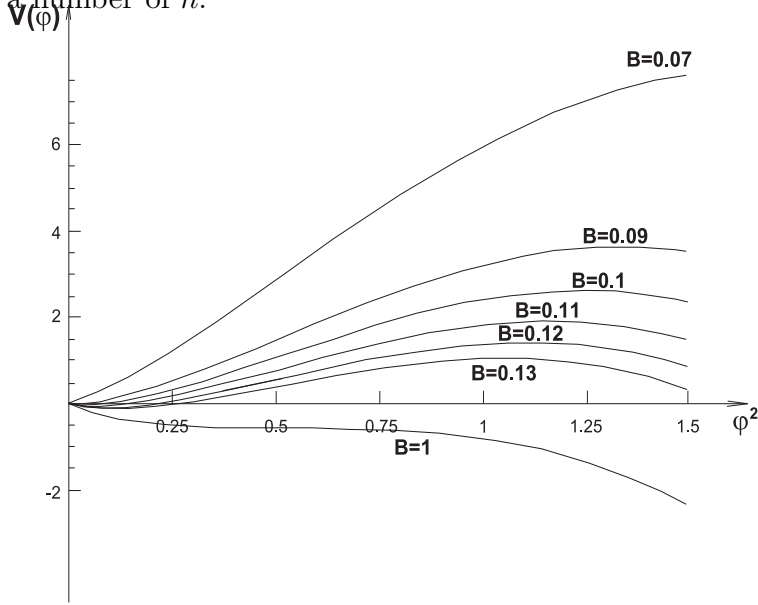}}
\end{picture}
\caption{Symmetry behaviour at high temperature and 'weak' magnetic
fields determined within the one-loop effective potential ${\cal V }= {\cal
V}^{(0)}+ {\cal V}^{(1)} $ for $K = 0.5$.}
\end{figure}
%                            Fig.8
\begin{figure}[hb]
\unitlength1cm
\begin{picture}(14,7)
\label{fd}
\put(0,-0){\epsfxsize=10cm\epsfysize=8cm\epsffile{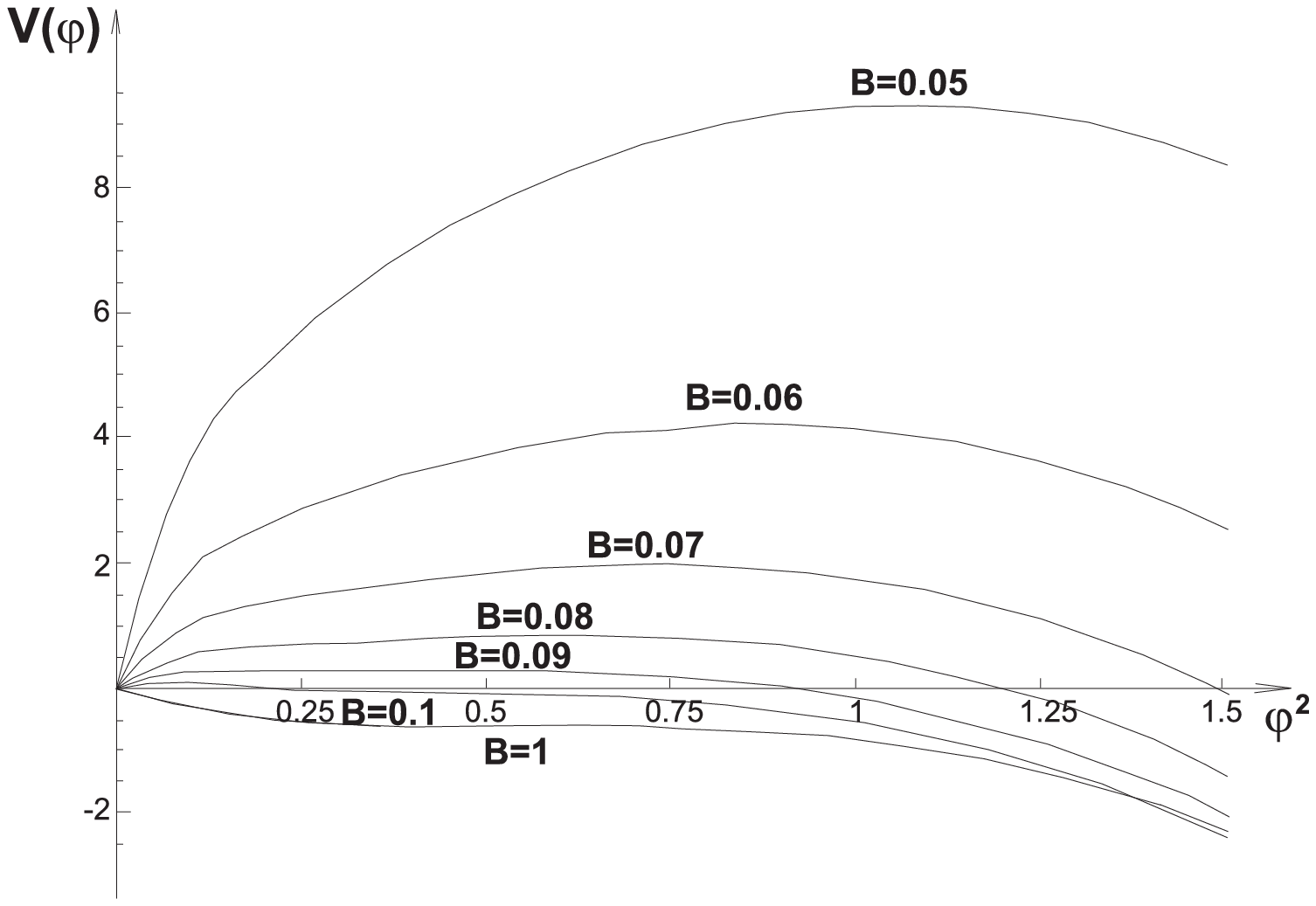}}
\end{picture}
\caption{Symmetry behaviour at high temperature and 'weak' magnetic
fields determined within the total effective potential $ {\cal V}=
{\cal V}^{(0)} + {\cal V}^{(1)}  + {\cal V}_{ring}$ for $K = 0.5$.}
\end{figure}
%                            Fig.9
\begin{figure}[hb]
\unitlength1cm
\begin{picture}(14,7)
\label{fd}
\put(0,-0){\epsfxsize=10cm\epsfysize=8cm\epsffile{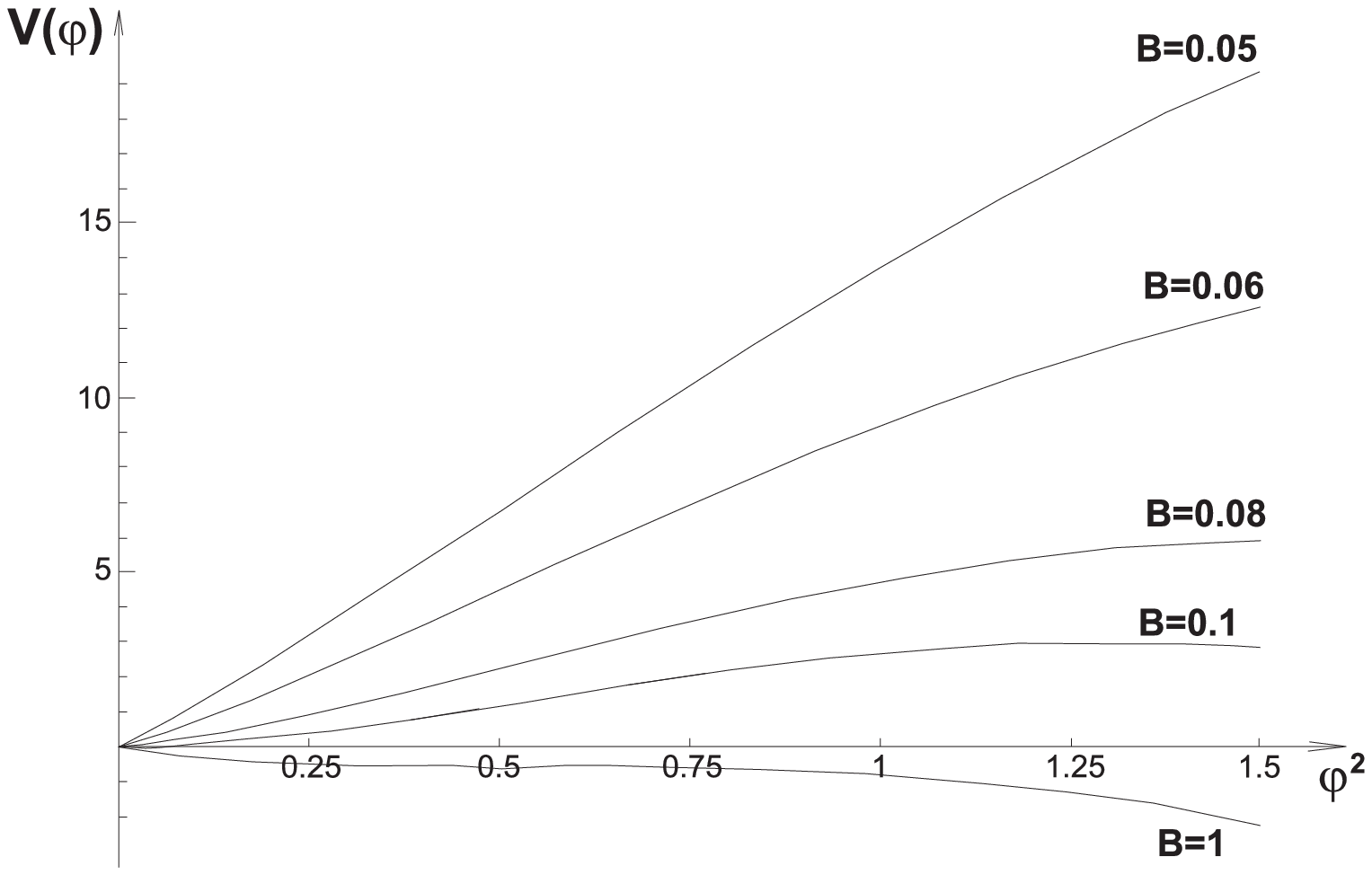}}
\end{picture}
\caption{Symmetry behaviour at high temperature and 'weak' magnetic
fields determined within the total effective potential $ {\cal V}=
{\cal V}^{(0)} + {\cal V}^{(1)}  + {\cal V}_{ring}$ for $K = 0.75$.}
\end{figure}
%                            Fig.10
\begin{figure}[hb]
\unitlength1cm
\begin{picture}(14,7)
\label{fd}
\put(0,-0){\epsfxsize=10cm\epsfysize=8cm\epsffile{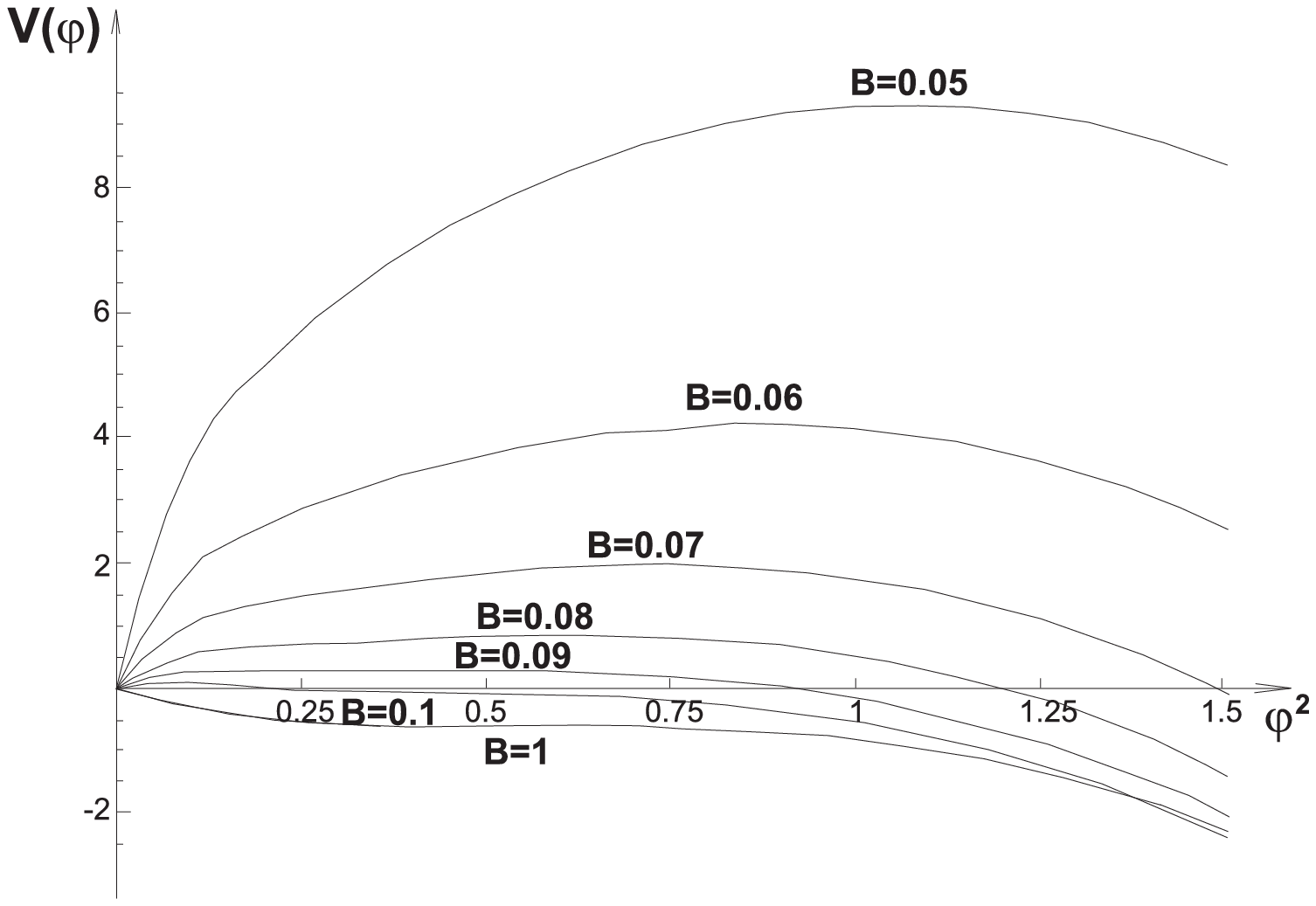}}
\end{picture}
\caption{Symmetry behaviour at high temperature and 'weak' magnetic
fields determined within the total effective potential $ {\cal V}=
{\cal V}^{(0)} + {\cal V}^{(1)}  + {\cal V}_{ring}$ for $K = 0.75$.}
\end{figure}
\end{document}